\def\year{2022}\relax
\documentclass[letterpaper]{article} 
\usepackage{aaai22}  
\usepackage{times}  
\usepackage{helvet}  
\usepackage{courier}  
\usepackage[hyphens]{url}  
\usepackage{graphicx} 
\usepackage{natbib}  
\usepackage{caption} 
\DeclareCaptionStyle{ruled}{labelfont=normalfont,labelsep=colon,strut=off} 
\usepackage{algorithm}
\usepackage{algorithmic}

\usepackage{newfloat}
\usepackage{listings}
\floatstyle{ruled}
\newfloat{listing}{tb}{lst}{}
\floatname{listing}{Listing}
\usepackage{subcaption}
\usepackage{changepage}

\title{Using Semantic Similarity and Text Embedding to Measure the Social Media Echo of Strategic Communications}
\author{Tristan J.B. Cann, Ben Dennes, Travis Coan, Saffron O'Neill and Hywel T.P. Williams
}
\affiliations{University of Exeter

}

\begin{document}

\maketitle

\begin{abstract}
Online discourse covers a wide range of topics and many actors tailor their content to impact online discussions through carefully crafted messages and targeted campaigns. Yet the scale and diversity of online media content make it difficult to evaluate the impact of a particular message. In this paper, we present a new technique that leverages semantic similarity to quantify the change in the discussion after a particular message has been published. We use a set of press releases from environmental organisations and tweets from the climate change debate to show that our novel approach reveals a heavy-tailed distribution of response in online discourse to strategic communications.
\end{abstract}

\section{Introduction}

Social media have exerted profound changes on established hierarchies of communication~\cite{Pearce19}. Diverse actors such as advertisers, charities, celebrities and governments seek to exert influence on the content and flow of information to promote their perspectives, products and agendas. In recent years, these attempts have seen increased awareness given their impact on democratic processes~\cite{Haenschen19} but precise measurement of the impact of these strategic communications campaigns on public discourse remains a challenge given the difficulty of defining suitable metrics and the volume of media content now available.

Typically, attempts to understand the response to media interventions make use of survey methods. \cite{Zhang18} surveyed individuals for their responsiveness to climate change information and found that exposure had a greater effect in more conservative areas. \cite{Feezell18} explored the impact of political content on social media, and found that not only did exposure increase awareness, it also reduced the rate of attention decay. \cite{Goldberg21} tested the response to banner ads on attitudes to climate change in the US, and found that the response was stronger among Republicans. \cite{Nyhan22} enhanced these findings by comparing the effects of factual, partisan and opinion pieces on climate change issue awareness. While individual attitudes were affected by each type of content, the short-lived gains from factual content were able to be quickly eroded by sceptical pieces. Survey-based experiments do present advantages by controlling the level of exposure that participants receive, but they are necessarily incapable of controlling all aspects of content exposure through longitudinal studies. Despite this, the impact of survey effects (and in particular social desirability bias in polarised contexts~\cite{Nederhof85}) mean that scholarly efforts would be enhanced by tools to study message effectiveness in real-world settings.

The rapid growth of digital media presents challenges for the scalability of analysis of media dynamics, including the measurement of influence. The volume of online news and social media content necessitates computational methods, but these have typically lacked the qualitative skill of survey methods. One potential avenue for analysing strategic messaging campaigns is using machine learning techniques to identify and measure influence in social media campaigns. \cite{Alizadeh20} used labelled data from Chinese, Russian and Venezuelan accounts to train a classifier for accounts linked to influence operations over both Twitter and Reddit posts using a range of content and metadata features. Their efforts were successful in specific contexts, but they note that the evolution of tactics by malicious actors hampers efforts to find a general set of features common to members of influence operations. Other attempts have investigated whether semantically coherent topics can be developed using content types on Twitter, such as that of \cite{JafariAsbagh14} who developed an efficient process capable of identifying organic trends in real time since they may indicate influence. Complementary to attempts to understand popular topics is the identification of individual user accounts that drive these trends. Previous efforts have considered user metadata and posting habits online~\cite{Azcorra18} or social network position~\cite{Hu18} to predict user influence. Most such efforts are hampered, however, by weak definitions (that use a partial substitute or proxy for influence) or incomplete data (since most data samples do not capture all interactions between individuals or content exposure due to practicalities and privacy constraints). Thus the challenge of measuring influence remains unsolved and new efforts are required to offer a more complete picture of interpersonal and organisational influence in online spaces.

Attempts to quantify information contagion or influence campaigns on contact networks have seen success. Some expand on the ideas of influence prediction discussed previously to give an estimate of the level of activity a particular topic will receive, rather than a binary indicator for whether the campaign was successful (e.g.~\cite{Weng14}). Many studies of information diffusion online rely on the use of specific parameters such as users, keywords, hashtags and URLs (see e.g.~\cite{Gabielkov16,Alperin19,Alizadeh20,Cruickshank20}) but these necessarily miss relevant discussions that do not use the predetermined search parameters.

Work by \cite{Bollenbacher22} is of particular relevance to the aims of this manuscript. Their research presented a means of quantifying the flow of attention between different conversational formats, using averaged word co-occurrence of texts to measure the change in agreement after the publication of a reference text. Regression analysis revealed that a small, but significant, flow of words emerged from news articles shared on Facebook to speeches made by MPs in the UK Houses of Parliament. Moreover, there was variation in the responses of different political parties with Labour politicians being more likely to use languange from news articles and social media interactions. This work has two main shortcomings, however. Using word co-occurrence will underestimate similarity when the two texts express the same message but with a different vocabulary. Additionally, although the authors note the short-lived engagement with online content~\cite{Gabielkov16}, they still choose to consider a two week post-event period, which may lead to any shorter term signals becoming lost.

In this paper we present a new framework for quantifying the uptake of strategic organisational messaging on social media. Our semantic similarity-based approach improves upon previous competitor techniques by adding robustness under changes to distinct but contextually similar vocabularies. We also present an alternative method that accounts for variation in background activity rates for effective comparison of messages released at different stages of the baseline activity cycle. To demonstrate the effectiveness of our metrics, we collect press releases from ten environmentally-active organisations and analyse their ``echo'' within the Twitter discussion of climate change. We find that our new metric allows for differentiation between messages and provides valuable context for understanding their impact.

The rest of this paper proceeds as follows. In the next section we present the conceptual framework we use to investigate influence. Later, we detail the development of our novel metrics, alongside the datasets used to test it. Finally, we detail the findings of our experiments before summarising our recommendations for future users.

\section{Conceptual model}\label{sec:conceptual_model}
From an engineering perspective, exerting influence on public debate can be conceptualised as the problem of transmitting information through a `noisy channel'. A source seeks to use communication to change some characteristic of a receiver; if successful, some change in the receiver state will be detectable and correlated with the communicated information. One possible example of this is a lobbying group using messaging to change public opinion. If their campaign is successful, then the target audience will, in some form, adopt aspects of the message. While the true desired effect might be an internal characteristic of the receiver, such as belief, opinion or feeling, these typically remain hidden. So to measure an effect we must focus on external characteristics, such as utterances or messages that reveal internal state.

\begin{figure}
    \centering
    \includegraphics[width=\linewidth]{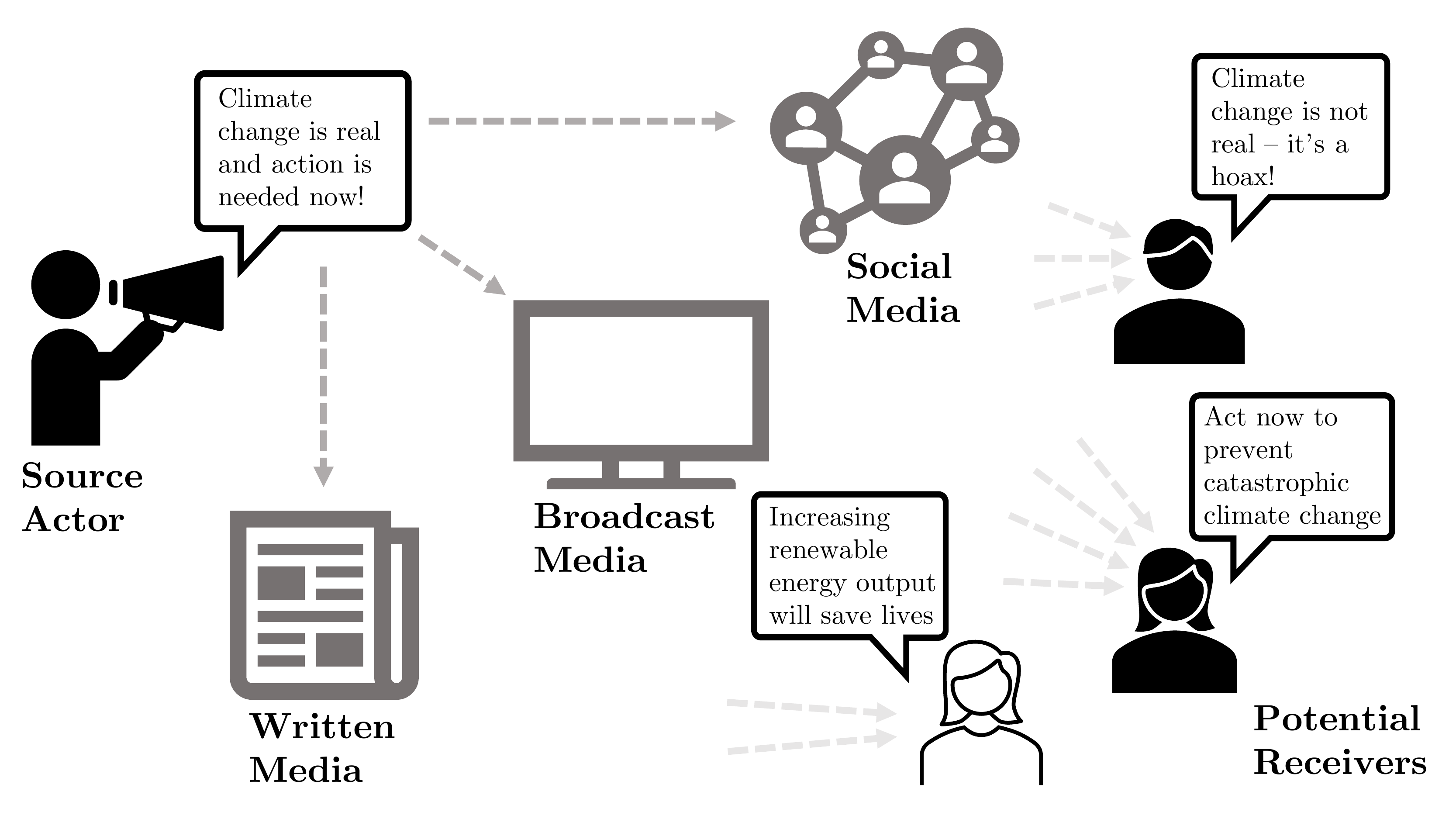}
    \caption{Illustrative schematic of our conceptual model. Strategic communications are published by groups and individuals who seek to shift the narrative in legacy or social media. How these messages filter through the media ecosystem and interpersonal influence is partially observable at best. The utterances of end users however can be observed and changes in message content can indicate they were potentially influenced by the strategic communication.}
    \label{fig:concept_schematic}
\end{figure}

A number of confounding factors add challenges to the types of inference that can be made about source influence. The primary complication is incomplete information about the transmission channel (we do not know the content to which the receiver was exposed, only the source message and receiver outputs) and the diversity of other potential influences (communications do not happen in isolation and we must assume that all receivers are exposed to many competing messages). As highlighted in Figure~\ref{fig:concept_schematic}, this leaves the problem as one of correlation, not causation; we cannot attribute causes, but can still seek to measure a correlated change in a receiver characteristic that occurs following a source communication. One final confounding factor in any analysis of communication channels is the impact of background media trends: any measure of a change in receiver behaviour should account for their `normal' behaviour from comparable periods.

As discussed in the previous section, much of the previous work in this area has used approaches that consider co-occurrence of keywords and tokens and other measures that are unaware of the context within which these words appear. Moreover, these methods are often reduced to considerations of frequency with limited consideration of the background trends. We propose taking advantage of the advances in the availability and scalability of text embedding techniques and semantic models to encode this context in our comparison of messages before and after source communications. By accounting for this change appropriately we can calculate receiver changes relative to the baseline media trends, and begin to reduce the impact of this confounding factor.

\section{Operationalising the model}\label{sec:developing_method}
In order to test the application of our conceptual model, we chose the context of strategic communications from environmental advocacy organisations sharing messages around climate change and seek to understand how they impact the online discussion of climate change on Twitter. This study makes use of two forms of text data: press releases from environmentally-focused organisations with a global reach and tweets about climate change, both for a two year period between 2019-11-01 and 2021-10-31.

The organisations were selected from a list of environmental organisations used in previous work~\cite{Weaver23}. From this list, we selected the 10 groups with the largest following on their main Twitter accounts. Table~\ref{tab:dataset_details} gives details of the included organisations and their reach in terms of Twitter followers and press releases published. For each organisation, a web crawler was constructed to parse the web page listing statements published as press releases by the organisations and extracting the text content within. In total 4,334 press releases were collected across the 10 organisations, and our analysis focuses on the 4,221 press releases with a full week of tweets before and after their release date (i.e. those between 8/11/19 and 24/10/21).

Tweets about climate change were collected via the Twitter streaming API by searching for the keywords \emph{climate change} and \emph{global warming}. The majority of the collection was completed in real-time, but short periods of interruption were recovered using historic searches through the V2 Academic API. While there are necessarily inconsistencies between the collection methods (e.g. the removal of posts in violation of the terms of service before historic collection), the volume of posts in our dataset within a given interval is comparable between the two collection methods. In total our dataset includes 67,672,904 tweets.

\begin{table*}
    \centering
    \begin{tabular}{|c|c|c|c|c|}
    \hline
        Organisation & Twitter handle & Followers & Press releases\\
    \hline
        World Wildlife Fund & @wwf & 3,936,636
 & 135\\
        Greenpeace & @greenpeace & 1,908,122
 & 135 \\
        Oxfam & @oxfam & 838,981
 & 335 \\
        Avaaz & @avaaz & 728,685
 & 17 \\
        Climate Reality Project & @climatereality & 611,028
 & 73 \\
        National Wildlife Federation & @nwf & 602,290
 & 503\\
        Public Citizen & @public\_citizen & 526,268
 & 775 \\
        350 & @350 & 391,928
 & 335 \\
        Sierra Club & @sierraclub & 384,747
 & 1,363 \\
        Natural Resources Defense Council & @nrdc & 345,191
 & 550 \\
         \hline
    \end{tabular}
    \caption{Details for the 10 environmental organisations defining our sample of press releases. Followership reflects the values recorded during collection in May 2022.}
    \label{tab:dataset_details}
\end{table*}

\subsection{Measuring text similarity}
To identify the semantic similarity between two texts, the gold standard method is manual analysis and labelling by a subject matter expert. However, due to the large volume of texts involved in our study -- in addition to the desire for a generalisable method that can be applied to new texts at any time -- such methodology is impractical. Instead we utilised a pre-trained deep learning model to perform text embedding, representing a sequence of text as a vector and thereby encoding the meaning of the text such that ``semantically similar sentences are close in vector space''~\cite{reimers19}. 

The deep learning model used was a sentence transformer, a type of model trained on large collections of text (such as Wikipedia and collections of literature~\cite{bao_unilmv2_2020}) and fine-tuned with semantic similarity data~\cite{reimers19} to improve the quality of text representations for the purpose of pairwise comparison. The sentence transformer used was \textit{all-MiniLM-L6-v2}~\cite{reimersmodel21}, chosen for its performance and efficiency. As outlined in the documentation of this model, the cosine similarity of vectors is an appropriate measure for similarity~\cite{sentence-transformers-docs, reimersmodel21}.

The \textit{all-MiniLM-L6-v2} model is designed to take short texts (less than 256 word pieces~\cite{reimersmodel21}, ``subword units [which] retain linguistic meaning''~\cite{song_fast_2021}) as input. As press releases are typically longer than this limit, the embeddings of all constituent sentences of a press release were averaged (after sentencisation with spaCy~\cite{spacy} following removal of extra white space characters) to result in an overall document embedding. The only pre-processing applied to the tweet text was the removal of all \emph{t.co} (Twitter-shortened) URLs due to their ubiquity in the texts and lack of linguistic information at face value.

The choice of similarity threshold at which a human reader would consider two texts to be similar is not an obvious one. Moreover in the context of strategic messaging by NGOs it is important to distinguish between tweets that cover the same topic and those that discuss the same aspects within a topic (such as references to a particular event or advocating a particular course of action). 

To test this, and validate our use of an embedding model over competing techniques, we prepared a sample of pairs of tweets and press releases for labelling by two human coders under a binary scheme where texts are labelled as similar under the stricter condition of requiring more specificity than the general topic. Under this scheme consider the following (fictional) press release:

\begin{adjustwidth}{0.3cm}{0.3cm}
\emph{Unprecedented warming in the Arctic, warmest February on record}

\noindent Temperature records from this year show that the average monthly temperature in the Arctic circle set a new record for the month of February. As global heating intensifies, such extremes will continue and new records set with increasing frequency. Global action is required to prevent a dangerous tipping point from being reached - contact your local representatives using the templates below.
\end{adjustwidth}
With reference to this press release, the tweet ``\emph{Sunny first day on my expedition in the Arctic}'' would be rated as dissimilar since it misses the key factor of extreme temperatures as a consequence of climate change. The tweet ``\emph{OMG, can't believe how extreme global heating is! Warmest Feb on record in Arctic, poor polar bears}'' would be rated as similar as it references both the extreme heat and geographic location.

For each of the press releases in our dataset we found their temporally relevant tweets (those falling within a week before or after the publication date of the press release) and calculated the cosine similarity between each pair of embeddings and from this set found press releases with which a range of similarity values was observed in the set of temporally relevant tweets. A number of press releases had very few (or even no) tweets with a similarity of greater than 0.7. However, as preliminary experiments had suggested that a suitable similarity threshold was at least this high, including such press releases would not give any useful information for calculating this threshold. Therefore, in order to filter such press releases out, the number of tweets for each press release with embedding similarity falling in 0.05-width bins between 0.5 and 0.8 were counted. Any press release without at least 4 unique tweet texts in each bin was excluded from sampling.

From the remaining press releases, four were sampled from each publishing organisation, and for each, four tweets were sampled from each similarity bin for a total of 960 pairs. These were shuffled and provided to two human coders without the similarity information, who discussed any disagreement in their labels to reach an agreement.

This labelled dataset allowed for comparison against two common methods for text similarity to assess their utility for the purposes of evaluating strategic communication campaigns. Jaccard similarity is the simplest of these, measuring the proportion of tokens appearing in either of two reference texts that appear in both. TF-IDF similarity is instead based on the frequency of tokens relative to their appearance across all texts in the corpus. Terms common in many of the texts have their weights reduced and consequently TF-IDF can be thought of as highlighting the characteristic terms for each text in turn.


One major area in which the TF-IDF and Jaccard similarity methods differ from semantic similarity is in their need for careful text pre-processing. We used sklearn~\cite{scikit-learn} to handle tokenisation and English stop word removal, alongside Porter stemming from NLTK~\cite{nltk} to allow optimal matching with variation in phrasing. Although the TF-IDF and Jaccard similarity metrics have no limitation in the lengths of text they are applied to, we use the same average sentence similarity as in the text embedding case for consistency. Investigation of alternative approaches using TF-IDF and Jaccard similarity measures saw qualitatively similar trends in similarity across the labelled testing set.

To identify the preferred similarity threshold required to distinguish similar and disssimilar pairs of texts we use the similarity scores found by text embedding, TF-IDF and Jaccard methods for pairs in our labelled set to define a binary classification problem. We used a range of values in the interval [0,1] as candidates and for each calculate the accuracy, F1 score and adjusted Rand index with respect to our human annotations for similarity.

\begin{figure*}
    \centering
    \begin{subfigure}[b]{0.45\linewidth}
    \includegraphics[width=\linewidth]{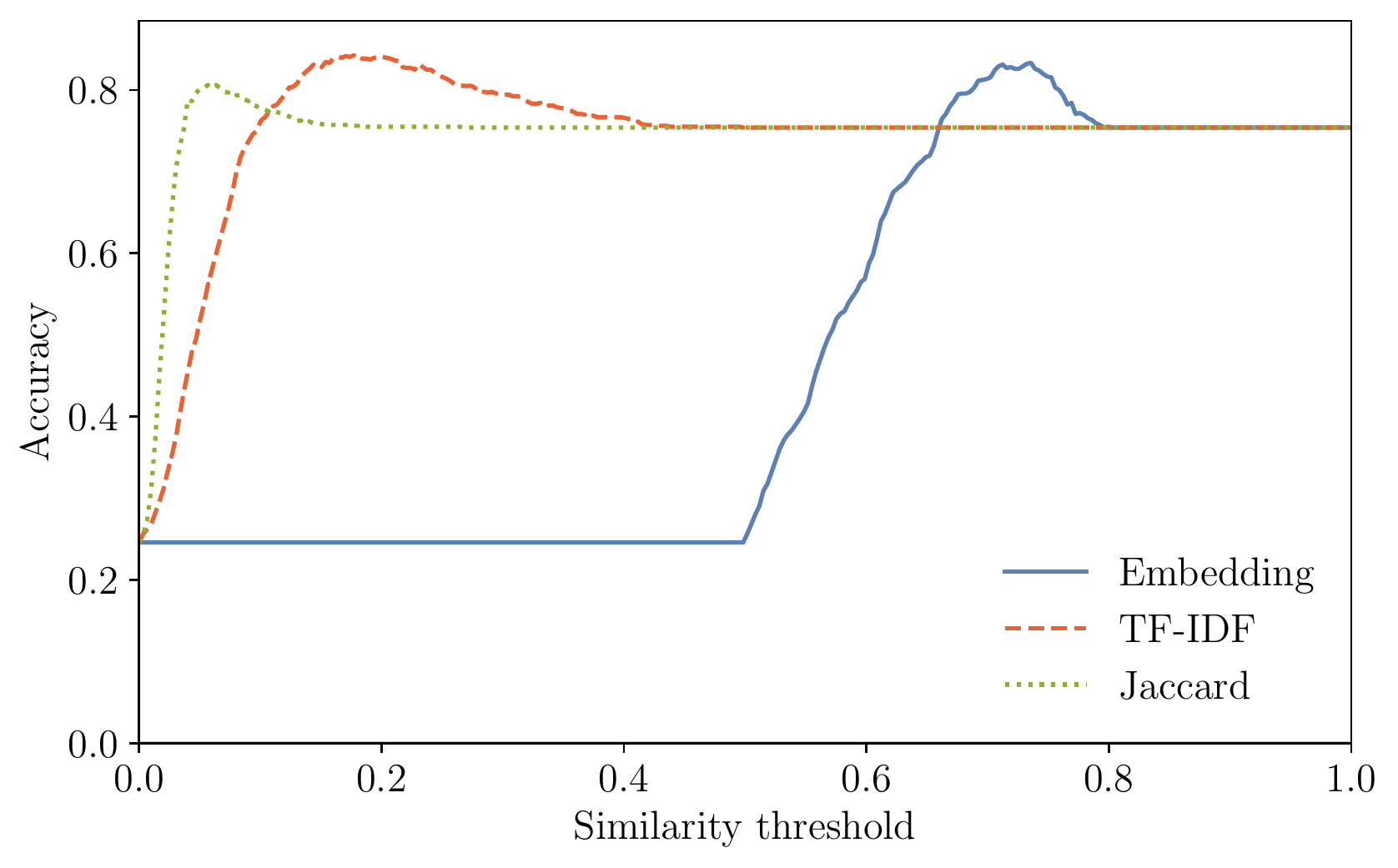}
    \caption{Accuracy}
    \label{fig:similarity_accuracy}
    \end{subfigure}
    \hfill
    \begin{subfigure}[b]{0.45\linewidth}
    \includegraphics[width=\linewidth]{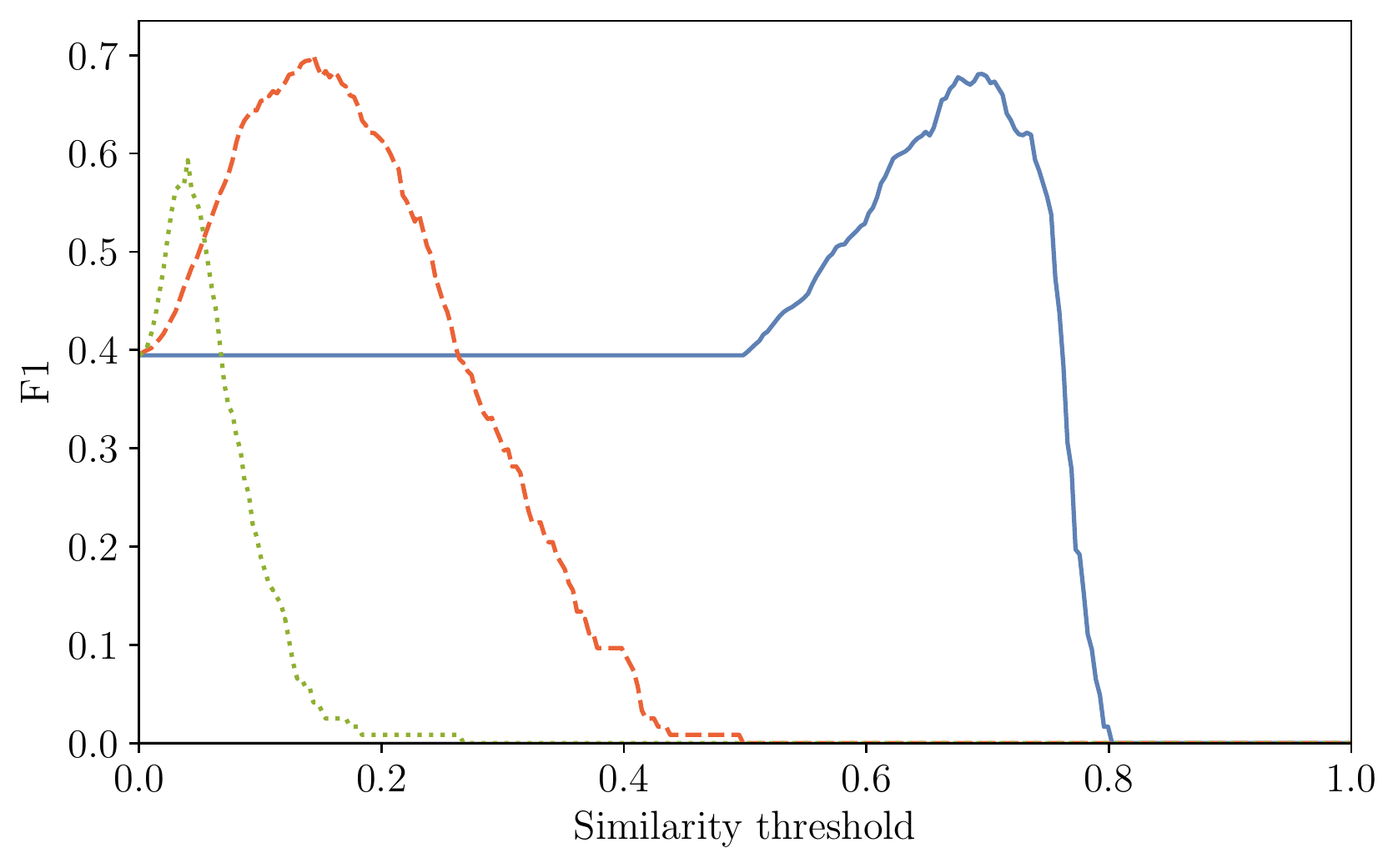}
    \caption{F1 score}
    \label{fig:similarity_f1}
    \end{subfigure}
    \begin{subfigure}[b]{0.45\linewidth}
    \includegraphics[width=\linewidth]{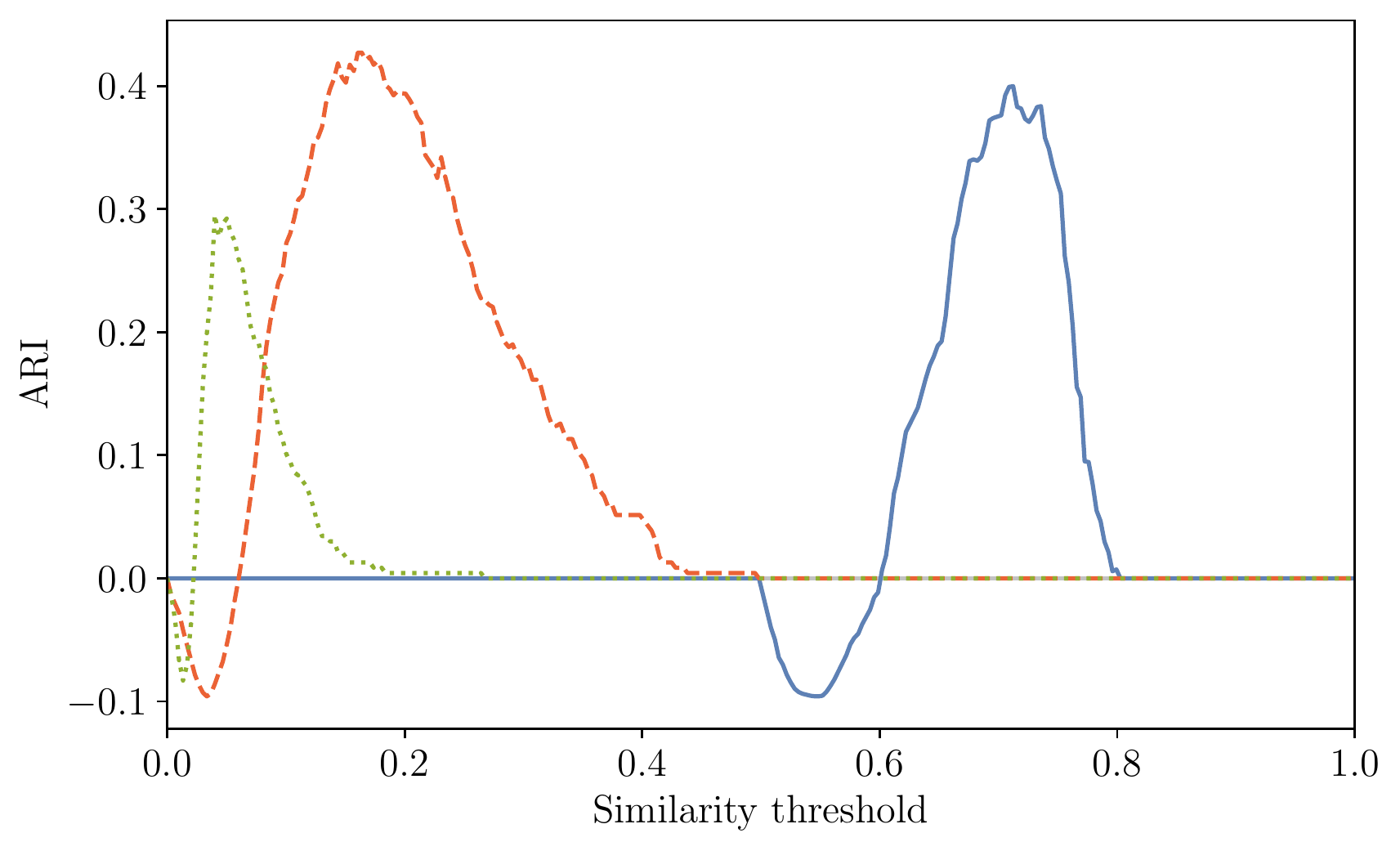}
    \caption{Adjusted Rand index}
    \label{fig:similarity_ari}
    \end{subfigure}
    \caption{Effectiveness of different similarity thresholds at discriminating between similar and dissimilar texts.}
    \label{fig:threshold_validation}
\end{figure*}

The results from our analysis are shown in Figure~\ref{fig:threshold_validation}. We note that given our sampling procedure for the pairs of tweets and press releases, the range of similarity values under text embedding is [0.5,0.8]. In regions outside of this interval, each of the evaluation statistics is unchanged for the embedding similarity as the threshold varies. This behaviour is repeated in the competitor metrics beyond their own ranges of similarity values. The first pattern revealed in Figure~\ref{fig:threshold_validation} is the comparatively low similarity values observed by the TF-IDF and Jaccard metrics, which reach maxima of 0.495 and 0.264 respectively. Despite this shift in the similarity values observed, we can see that each of the metrics was able to distinguish between similar and dissimilar pairs more accurately than labelling by the most-common class (dissimilar) as evidenced by Figure~\ref{fig:similarity_accuracy}. Moreover, running pairwise comparisons of the labels assigned using threshold values in their respective preferred regions showed good agreement across approaches: each pairwise accuracy was greater than 0.7 using a threshold of 0.7 for embedding similarity, 0.2 for TF-IDF similarity and 0.08 for Jaccard similarity. Figs.~\ref{fig:similarity_f1} and~\ref{fig:similarity_ari} suggest that the Jaccard similarity metric performs the worst when accounting for the imbalanced classes in the labelled data (approximately 25\% were rated as similar by our human coders). We expect that these shortcomings in the TF-IDF and Jaccard similarities come from the challenges of our particular test case. Tweets are shorter pieces of text given their character limits, and therefore can only include a limited number of words. Since the TF-IDF and Jaccard metrics rely only on presence of tokens and not the context within which those tokens appear, they are unable to capture the maximum amount of information from the limited text available in tweets. As such, our choice of using text embedding as a preferred metric is validated in our context.

From Figure~\ref{fig:threshold_validation} we can see that there is consistently peaked behaviour under each of the similarity metrics and each of the reported benchmarking statistics. Focusing now on the results for the text embedding case, we can see that the maximum value observed for accuracy, F1 score and adjusted Rand index is found close to a similarity threshold of 0.7 (accuracy at 0.736, F1 score at 0.696, adjusted Rand index at 0.712). In the context of this consistency, we choose 0.7 as an appropriate threshold to recognise texts as `sufficiently similar' for our purposes.

\subsection{Measuring the echo}
We now detail the method for calculating the echo of a given press release, for which we require the press release text and its release date. Given the release date, we collect all tweets in our dataset within a predetermined window around the release date (including the release date itself). 

The press release and tweet texts are each embedded into a 384-dimensional vector space wherein we can calculate the semantic alignment between texts using cosine similarity. Using the preferred threshold of 0.7, identified in the previous section, we can then label tweets as ``sufficiently similar'' if the cosine similarity of their embedding and the press release embedding is greater than this threshold. 

Our first proposed echo metric considers the absolute change in sufficiently similar tweets. We establish the baseline topic visibility by calculating the average number of sufficiently similar tweets on the days preceding press release publication and similarly measure the publication impact by calculating the average number of sufficiently similar tweets on the days after publication (typically including the release date in this window). The metric, which we refer to as $\Delta_{raw}$, is finally calculated as the difference between the post-day calculation and the pre-day calculation. In other words,
\begin{equation}
    \Delta_{raw} = \frac{1}{|\mathrm{Post}|}\sum_{i \in \mathrm{Post}} |S_i| - \frac{1}{|\mathrm{Pre}|}\sum_{j \in \mathrm{Pre}} |S_j|,
\end{equation}
where $S_i$ is the set of sufficiently similar tweets on day $i$, and Pre and Post are the sets of pre- and post-release days considered respectively. Positive values of $\Delta_{raw}$ indicate more sufficiently similar tweets after the publication date, negative values indicate fewer sufficiently similar tweets after publication and values around zero indicate little change overall. 

Social media activity is inherently variable on a day-to-day basis, and as such we present a complementary metric called $\Delta_{prop}$ to normalise for this variation. Where $\Delta_{raw}$ considers the absolute number of tweets, $\Delta_{prop}$ considers the rate of sufficiently similar tweets relative to daily changes in background activity, that is
\begin{equation}
    \Delta_{prop} = \frac{1}{|\mathrm{Post}|} \sum_{i \in \mathrm{Post}} \frac{|S_i|}{|T_i|} - \frac{1}{|\mathrm{Pre}|}\sum_{j \in \mathrm{Pre}} \frac{|S_j|}{|T_j|},
\end{equation}
where $T_i$ is the total number of tweets on day $i$.

\begin{figure*}
    \centering
    \begin{subfigure}[t]{0.45\linewidth}
    \includegraphics[width=\linewidth]{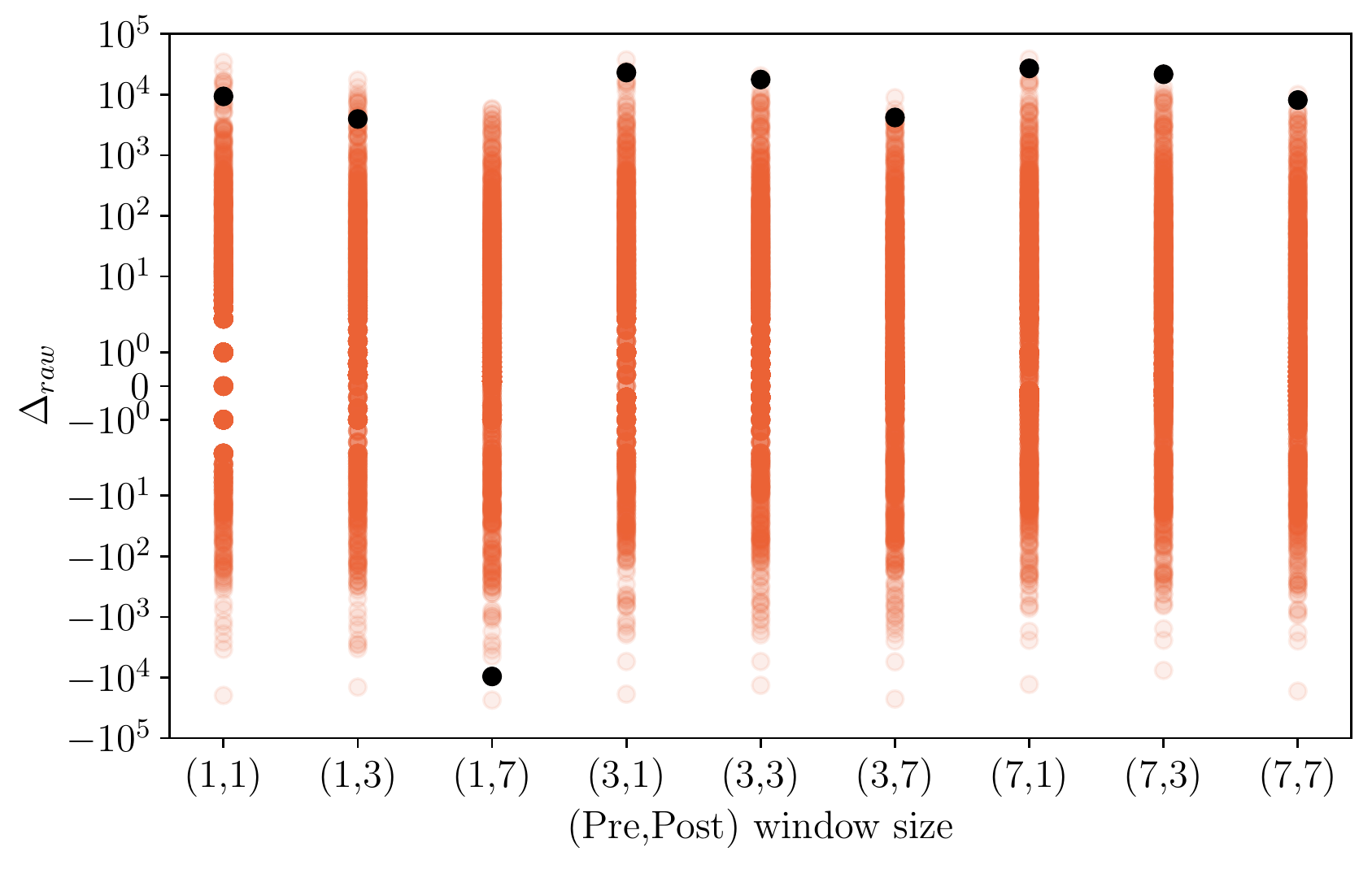}
    \caption{$\Delta_{raw}$ for a press release from 350 on 2020-02-18 discussing Jeff Bezos' \$10 billion pledge to create a fund fighting climate change.}
    \label{fig:max_raw_window_example}
    \end{subfigure} \hfill
    \begin{subfigure}[t]{0.45\linewidth}
    \includegraphics[width=\linewidth]{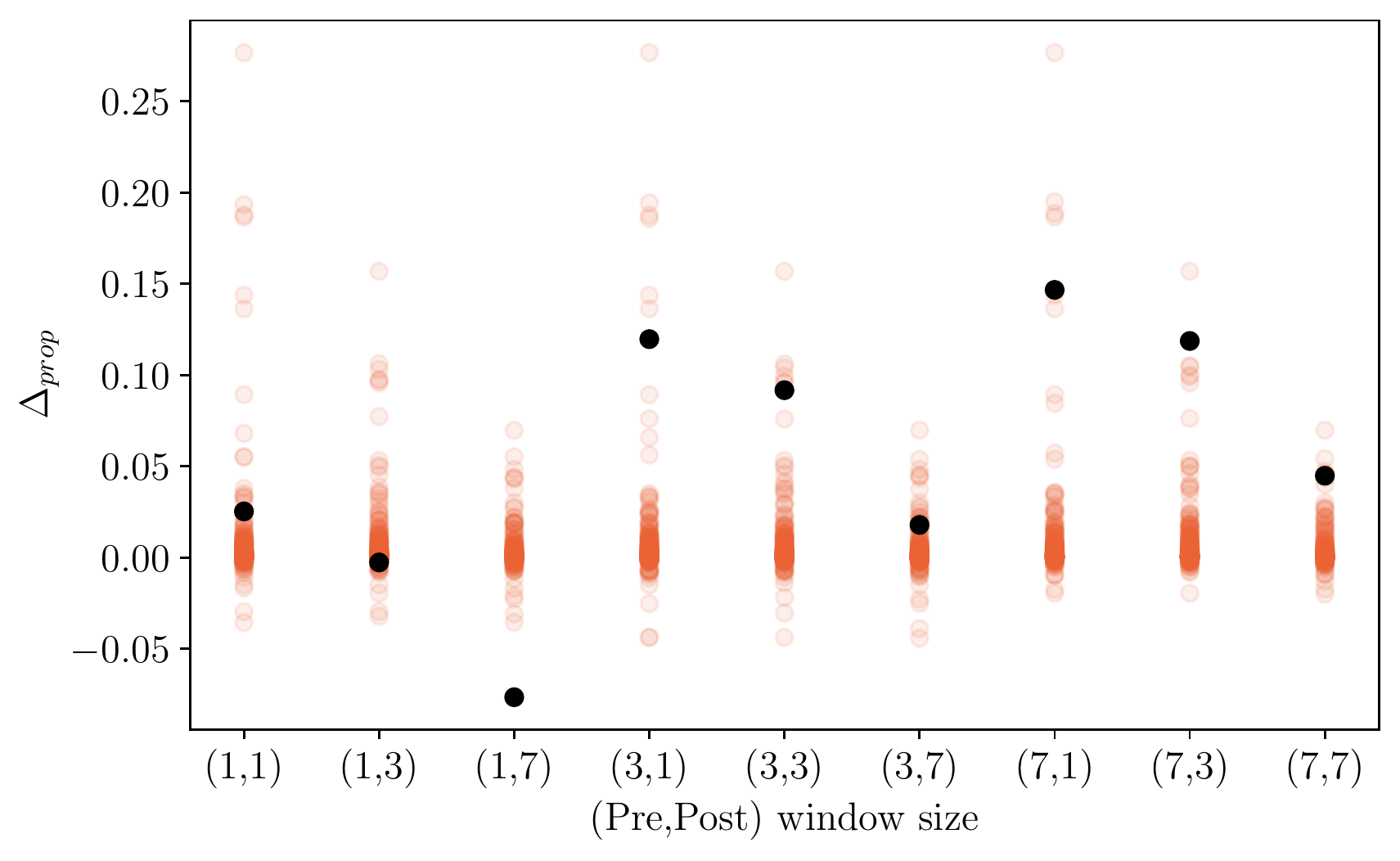}
    \caption{$\Delta_{prop}$ for a press release from 350 on 2020-02-18 discussing Jeff Bezos' \$10 billion pledge to create a fund fighting climate change.}
    \label{fig:max_prop_window_example}
    \end{subfigure}    
    \begin{subfigure}[t]{0.45\linewidth}
    \includegraphics[width=\linewidth]{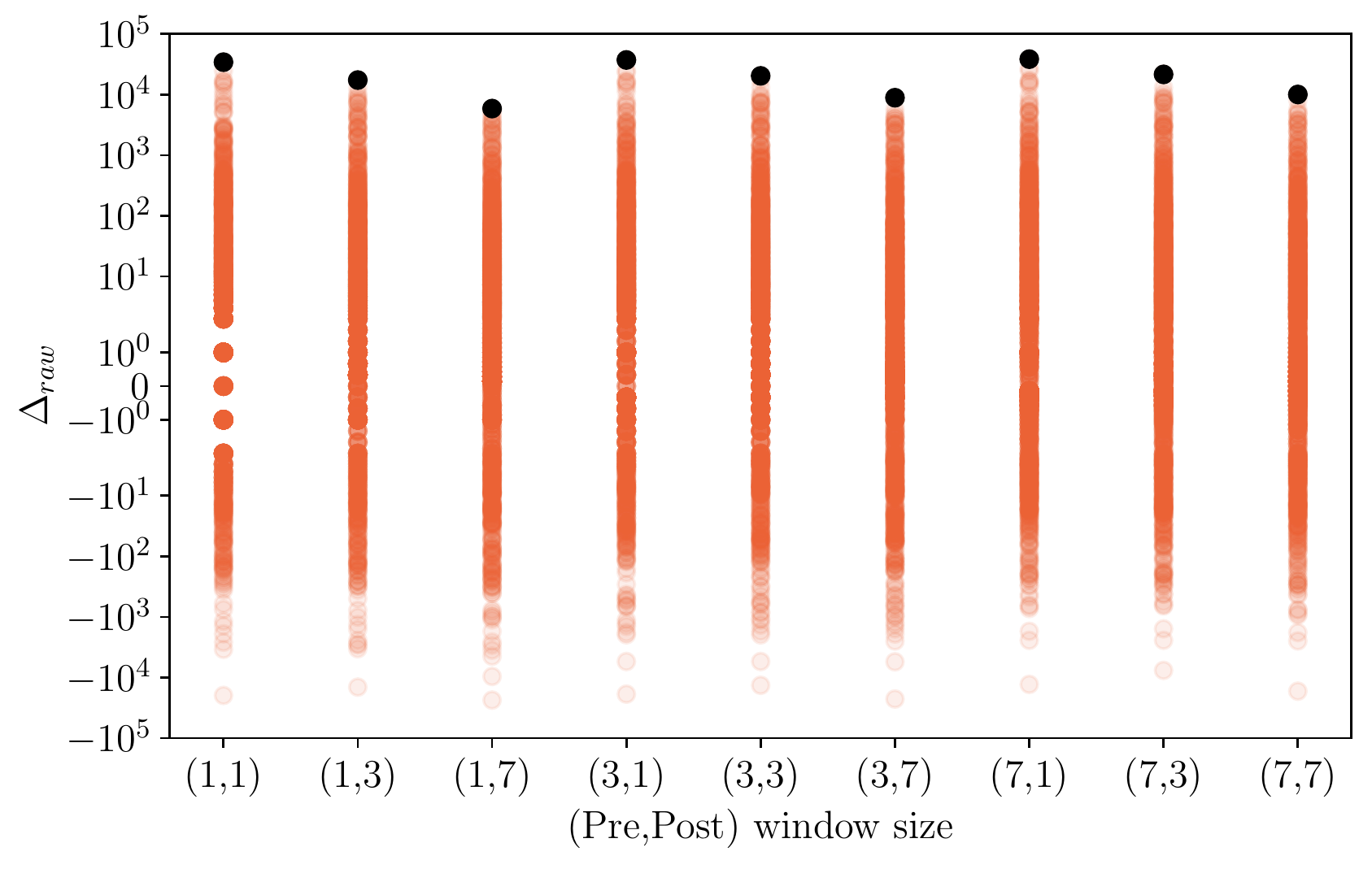}
    \caption{$\Delta_{raw}$ for a press release from Greenpeace on 2021-08-09 highlighting the need for urgent climate action following the publication of the latest IPCC WG1 report.}
    \label{fig:max_raw_window_example2}
    \end{subfigure} \hfill
    \begin{subfigure}[t]{0.45\linewidth}
    \includegraphics[width=\linewidth]{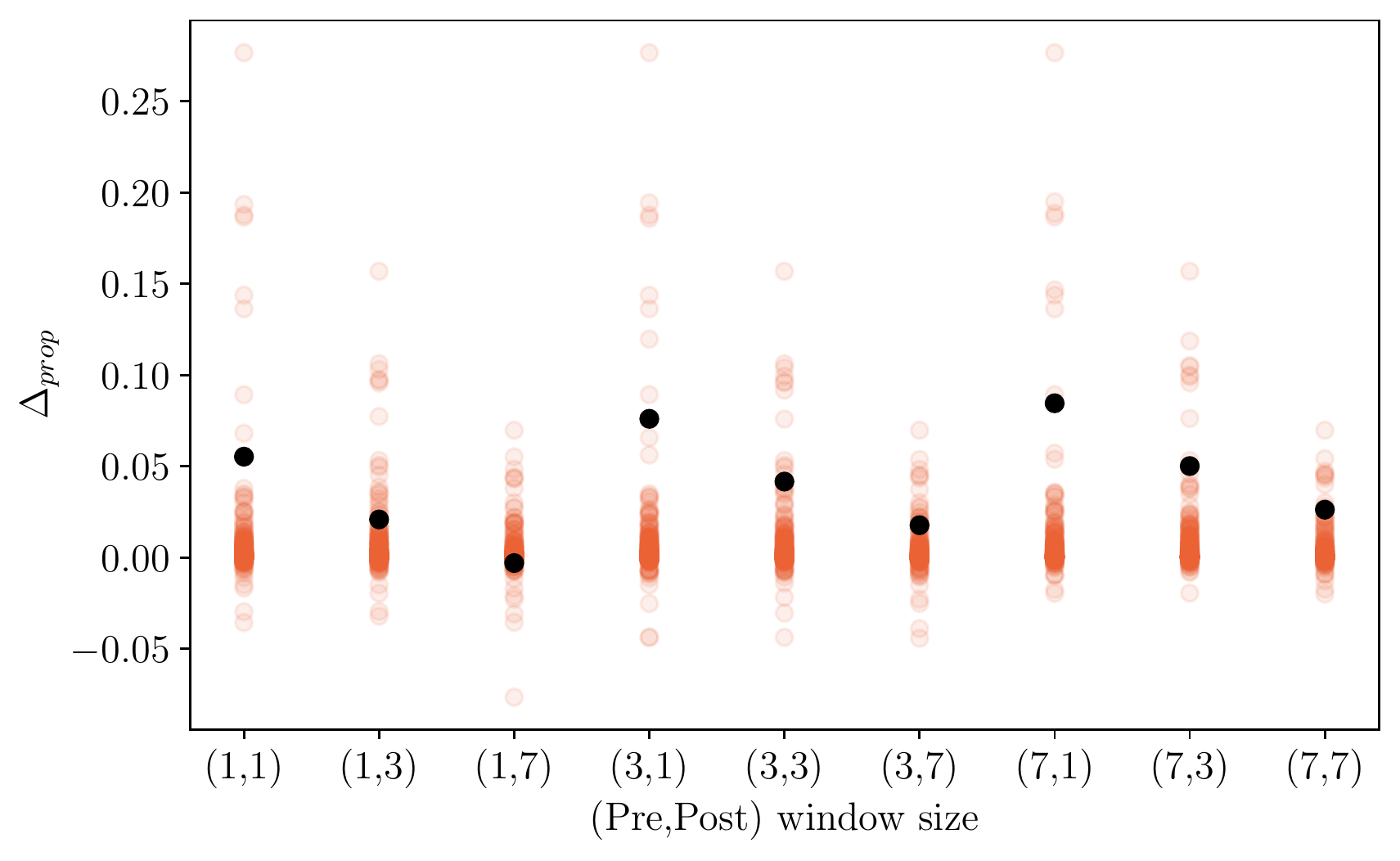}
    \caption{$\Delta_{prop}$ for a press release from Greenpeace on 2021-08-09 highlighting the need for urgent climate action following the publication of the latest IPCC WG1 report}
    \label{fig:max_prop_window_example2}
    \end{subfigure}  
    \caption{Selected examples from the press releases that score most highly by $\Delta_{raw}$. The black points in each panel show the position of the highlighted press release relative to all other press releases in our sample under the tested combinations of pre- and post-release window sizes.}
    \label{fig:diff_window_echo_examples}
\end{figure*}

The size of the pre- and post-release windows poses a question of which combination is preferable. We tested each combination of one, three and seven days before and after the release date on our sample of press releases from environmental organisations to determine the impact of these choices. In Figure~\ref{fig:diff_window_echo_examples} we show how our metrics perform under these conditions, and also highlight two of the most successful examples under $\Delta_{raw}$. We highlight a comment from 350 about Jeff Bezos' \$10 billion climate pledge in February 2020 in Figs.~\ref{fig:max_raw_window_example} and~\ref{fig:max_prop_window_example}. We see here that in most cases there is some variation in the echo scores returned for a given piece of text, but a successful message will still be considered as such in most combinations of pre- and post-release window size. The case of one day preceding release and seven days after release stands out as anomalous for this press release where it sees a large, negative echo in contrast to the other cases. This situation occurs due to a combination of a large amount of activity on the day before release coupled with short-lived activity after release resulting in the post-release window average being comparatively small. This example highlights an extreme case and illustrates the potential pitfalls with failing to establish a suitable baseline comparison for the post-release window and in observing beyond the lifetime of attention to the topic.

Figure~\ref{fig:diff_window_echo_examples} reveals another trend in the responses of our metrics to different window sizes when looking at the broader distribution trends. Primarily we can see that the size of the post-release window has a clear effect on the range of values observed, whereas the size of the pre-release window has much less impact. As the post-release window expands to include days further from the release, we see that the range of values observed across all press releases decreases, as does the echo score observed for the specific examples highlighted. These observations make intuitive sense in the context of media ecosystems and our conceptual model. In the case of a truly novel message transmitted in the press release, the number of sufficiently similar messages produced before this content is shared will be low and stable over time, therefore removing any dependence on pre-window size from our echo metrics. This is not always the case however since some messages can be retrospective (referencing events that occur in the past) or periodic (referencing events that occur with some frequency). Such cases would particularly benefit from a choice of pre-release window size that covers the range of relevant behavioural rhythms among individuals such as daily or weekly activity habits. The suitability of a choice for post-release window size has similar considerations around the lasting relevancy of a message. On social media, narratives in the media ecosystem evolves rapidly and quickly shift from issue to issue~\cite{Castillo14}. Typically however, most awareness of media issues is short-lived which means that shorter post-release windows should be preferred.

Given these observations we present the following recommendations for the choice of pre- and post-release window sizes. In all cases, we recommend avoiding the use of a single day in the pre- and post-release windows to limit the impact of anomalous days. Our context of politically relevant messages observed through the lens of social media gives a rapid turnover of events that can have relevance of several days and in this case recommend a choice of three days for the post window. Seven days is our preferred choice for the size of the pre-release window to fully capture weekly baseline trends and allow for large echo scores for reactive messages such as the example in Figure~\ref{fig:max_raw_window_example}.

\section{Interpreting the echo}\label{sec:echo_results}

\begin{figure*}
    \centering
    \includegraphics[width=0.9\linewidth]{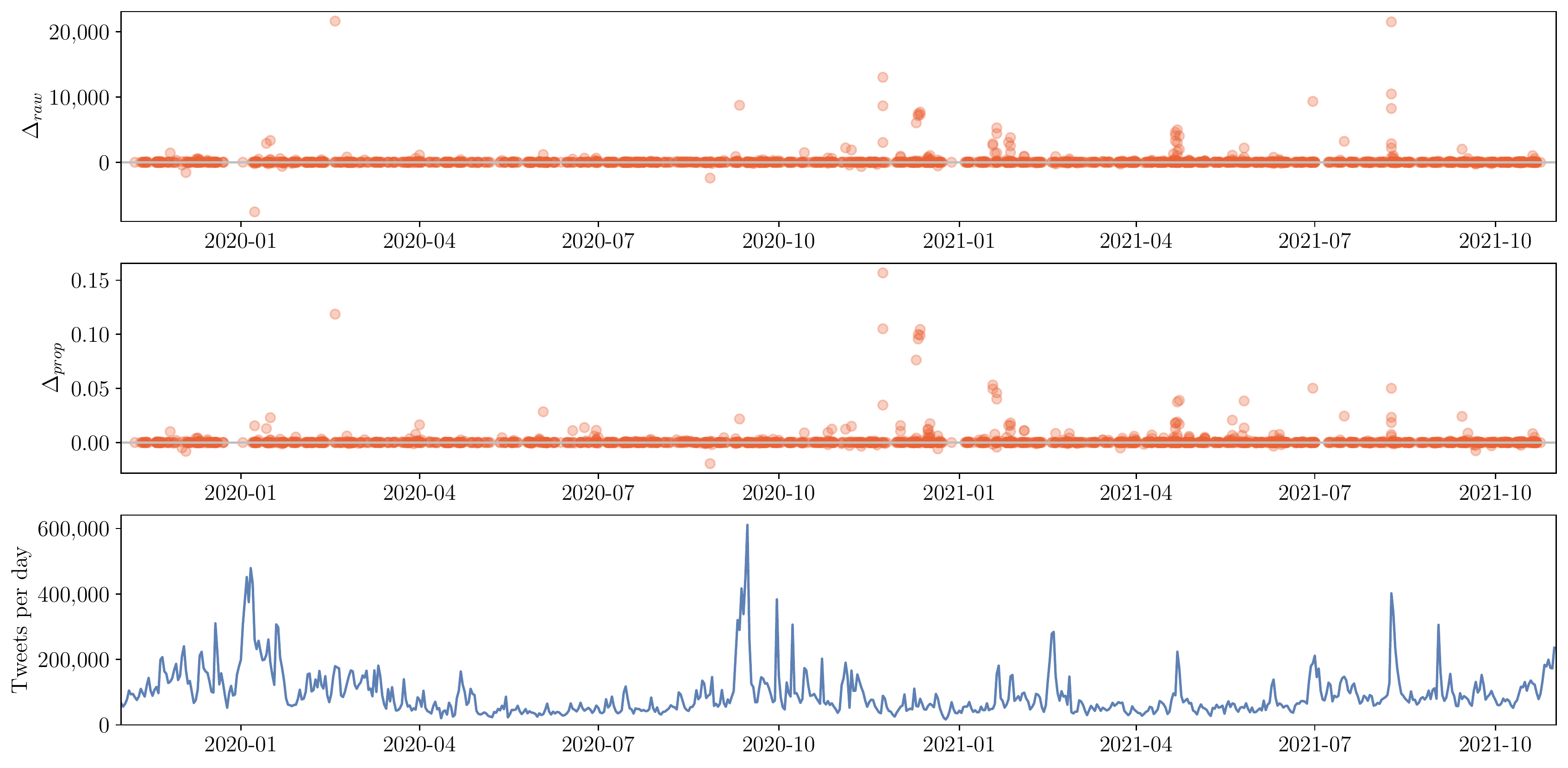}
    \caption{Scores for the echo metrics for all press releases in the dataset that had sufficiently similar tweets and data available for the seven days before and after the release date. The bottom panel plots the number of tweets collected each day for context.}
    \label{fig:all_pr_deltas}
\end{figure*}

We now explore the general behaviours of both $\Delta_{raw}$ and $\Delta_{prop}$ over the press releases in our dataset using the preferred parameter choices of a similarity threshold of 0.7, pre-release window size of seven days and post-release window size of three days, as shown in Figure~\ref{fig:all_pr_deltas}. Here we can see that the majority of press releases have very little impact on the Twitter conversation of climate change (of those with sufficiently similar tweets, 95th percentiles: $\Delta_{raw}$ 351.67, $\Delta_{prop}$ 0.0041) compared to the maximum values observed ($\Delta_{raw}$ 21,655.52, $\Delta_{prop}$ 0.157) and suggest a strongly positively skewed distributions. Some press releases have negative $\Delta$ values, which indicate that interest in the topic was greater before the publication date, but this is most apparent in the case of $\Delta_{raw}$ when a press release is referring to a recent natural disaster (the Australian bush fires in January 2020).

Across the set of all press releases studied a few patterns emerge. Firstly, more than 50\% of all texts (2,364/4,221) had no sufficiently similar tweets within the period defined by our pre- and post-release windows, that is their messages were not repeated on Twitter close to their release date and we omit such cases from all figures. We can also see that $\Delta$ values are not driven solely by baseline activity in the Twitter collection but do see some evidence of temporal clusters of influential communications. These typically arise from notable events such as John Kerry being named as US International Climate Envoy after the victory of President Biden in the 2020 US presidential election (2020-11-23), the release of the IPCC Sixth Assessment Report (2021-08-09) or the Leaders' Climate Summit on Earth Day (2021-04-22).

\begin{figure}
    \centering
    \includegraphics[width=\linewidth]{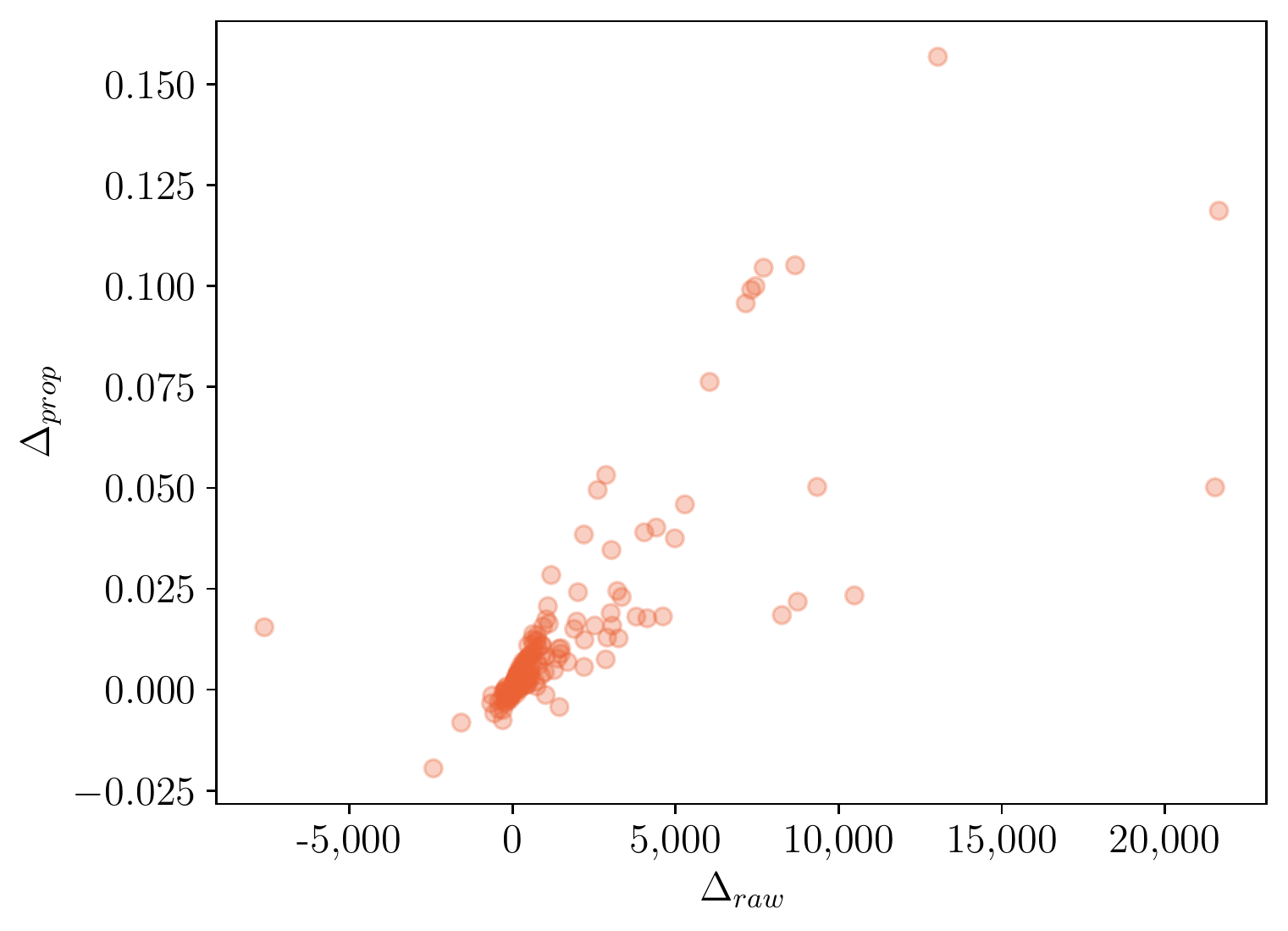}
    \caption{Scatter plot of $\Delta_{raw}$ against $\Delta_{prop}$. Correlation between the two metrics is strong (Pearson's $R=0.808,p<0.01$).}
    \label{fig:raw_prop_scatter}
\end{figure}

When comparing trends observed for $\Delta$ values in Figure~\ref{fig:all_pr_deltas}, the successful press releases under one metric tend to be similarly successful under the other. $\Delta_{prop}$ however is more successful at highlighting important communications in periods of lower baseline Twitter activity, given its inherent normalisation of background activity trends. Figure~\ref{fig:raw_prop_scatter} complements this by showing a clear correlation between $\Delta_{raw}$ and $\Delta_{prop}$ (Pearson's $R=0.808,p<0.01$). The two press releases with the largest $\Delta_{raw}$ (as shown in Table~\ref{tab:top_prs}) are outliers from this trend, as is the press release with the smallest $\Delta_{raw}$ (in which 350 discussed the catastrophic Australian bushfires in January 2020 and called on the Australian government to take action). These outliers highlight considerations to make when comparing our $\Delta$ metrics. Firstly, $\Delta_{raw}$ is affected by changes in the background rate of Twitter posts. The Australian bushfires example shows this clearly as a reactive message to a major event. Calculating $\Delta_{prop}$ for the same message we instead find a small positive effect, which shows that the large negative value of $\Delta_{raw}$ observed was linked to the decline in climate change activity on Twitter. A similar reason explains the outliers with high $\Delta_{raw}$. The IPCC Sixth Assessment WG1 report release is a major event in the climate change discussion on Twitter, producing a more than five-fold increase in daily posts compared to the previous week. Relevant press releases published towards the beginning of this spike therefore received large $\Delta_{raw}$ values.

We focus now on the most influential press releases as highlighted by our new metrics. Table~\ref{tab:top_prs} lists the five most successful by $\Delta_{raw}$ and $\Delta_{prop}$. A few patterns are revealed from these highly successful communications. Firstly, 350 and Climate Reality Project are well-represented and each contributes a message that is highly rated on both of our metric formulations: discussion of a philanthropic pledge by Amazon founder Jeff Bezos and the selection of John Kerry to a position of climate leadership by President-elect Biden respectively. More broadly, the most influential communications are tied into major events such as the incoming Biden administration, the IPCC Sixth Assessment Report publication and investigative reporting into the lobbying efforts of ExxonMobil.

\begin{table*}
    \centering
    \begin{tabular}{|c|c|p{6cm}|c|p{6cm}|}
        \hline Rank & $\Delta_{raw}$ & URL & $\Delta_{prop}$ & URL \\ \hline
        1 & 21,655.5 & 350.org/press-release/bezos-climate-pledge/ & 0.157 & climaterealityproject.org/press/statement-ken-berlin-secretary-john-kerry \\ \hline
        2 & 21,537.4 & greenpeace.org/international/press-release/49125/ipcc-report-signals-decisive-moment-for-humanity-urgent-climate-action-needed/ & 0.119 & 350.org/press-release/bezos-climate-pledge/ \\ \hline
        3 & 13,038.9 & climaterealityproject.org/press/statement-ken-berlin-secretary-john-kerry & 0.105 & 350.org/press-release/kerry-international-climate-envoy/ \\ \hline
        4 & 10,479.7 & 350.org/press-release/350-org-reacts-to-ipcc-report-on-the-state-of-our-climate/ & 0.105 & wwf.panda.org/wwf\_news/press\_releases/ ?1174191/Climate-Ambition-Summit-2020 \\ \hline
        5 & 9,340.3 & 350.org/press-release/exxon-tapes-climate-crimes/ & 0.100 & climaterealityproject.org/press/statement-former-vice-president-al-gore-fifth-anniversary-paris-climate-agreement \\ \hline
    \end{tabular}
    \caption{The five most influential press releases by each $\Delta$ metric.}
    \label{tab:top_prs}
\end{table*}

The interpretations of $\Delta_{raw}$ and $\Delta_{prop}$ in an experimental context vary. $\Delta_{raw}$ measures the average change in sufficiently similar pieces of content and therefore gives an indication of the size of the following gained by a particular message. Analysis of this figure should always consider the context of changing background activity rates, and comparisons between contexts should also be aware of any differences between their respective audience sizes. $\Delta_{prop}$ on the other hand gives a measure of change relative to the population of interest and therefore should be thought of as the change in the proportion of the whole reference conversation that is sufficiently similar to the chosen message. More specifically, a $\Delta_{prop}$ of 0.1 means that on average 10\% more of the conversation is similar to the chosen message after its release.

Given our example dataset and the interpretation of each metric we can now discuss how these metrics can be used to determine whether a particular message is successful. We observed highly-skewed distributions for both $\Delta_{raw}$ and $\Delta_{prop}$ where most of observed values are small relative to the maximum. In eight of the nine combinations of pre- and post-release window sizes for $\Delta_{prop}$, the 97th percentile is less than 0.01 which indicates that such a value is a consistent threshold for the most impactful communications. Identifying such a threshold for $\Delta_{raw}$ is more challenging given the inherent variability of the measure in different contexts. With reference to the Twitter conversation surrounding climate change the 97th percentile ranges from approximately 400 to approximately 1200 (which is not surprising given the patterns shown in Figs.~\ref{fig:max_prop_window_example} and~\ref{fig:max_prop_window_example2}). The lower end of this range is observed when using seven days post-release which is likely to include too much `business as usual' activity after attention to the message has faded. Excluding these cases, the lower range limit increases to approximately 850 and shows a similar approximately one percentage point change relative to average daily tweets in our sample (approximately 93,000). 

We can summarise the recommendations for our echo metrics as follows. $\Delta_{raw}$ and $\Delta_{prop}$ should be used as complementary metrics that capture absolute and relative change respectively. We recommend using a value of 0.7 as the threshold for semantic similarity for the best balance of relevance and coverage when comparing reference texts. The choice of pre-release window size was less impactful on the trends observed so we therefore recommend using seven days to account for daily and weekly activity rhythms. In the context of fast-moving social media discussions it is important to choose a shorter post-release window size that avoids including many days after attention has waned. As such, three days is a suitable choice for post-release window size. Finally, the distribution of $\Delta_{raw}$ and $\Delta_{prop}$ scores we observed suggest that echo scores equivalent to an average increase in coverage of one percentage point should be considered indicative of messages with the largest social media discussion.




\section{Discussion}\label{sec:discussion}
In this paper we introduced a new methodology for evaluating the impact of strategic communications. Our new metrics use text embedding to measure the change in semantic similarity between the text of the reference strategic message and a set of texts from an audience of interest close to its release date. Given the variability in posting behaviours over time and among groups, we presented two variants to measure the impact or `echo' of a press release: $\Delta_{raw}$ which measures the change in absolute activity and $\Delta_{prop}$, which measures the change in activity relative to the baseline activity rate. To test this approach, we focused on ten organisations promoting action on climate change and tweets to capture the conversational echo of organisational messaging over a two year period.

Our analysis of this new technique focused on understanding the responsiveness of Twitter users to messaging within the online debate around climate change. We found a highly skewed distribution of echo scores, with the majority of the press releases collected seeing very little change in the level of semantically similar posts in the discourse. However, we did observe several highly successful strategic communications with an increase of the order of ten thousand semantically similar posts after their release date. As a result, we can be confident that our approach is suitable for differentiating between content with varying levels of uptake in the public discourse on social media.

We also find some evidence of temporal co-occurrence in successful messaging. In Figure~\ref{fig:all_pr_deltas} we can see vertical bands of increased responsiveness to strategic communications. Many of these arise around events in the US political discourse such as the 2020 US presidential election and the initial actions by newly-elected President Biden. In some cases these are driven by major events promoting wider awareness of and participation in the climate change debate (e.g. the 2021 Earth Day Summit, the 2021 IPCC WG1 report release) but in others, the same events do not precipitate an increase in activity. This suggests that timing and consistency in messaging may be a key component in promoting wider success.

One important consideration arising from our $\Delta$ metrics is in the use of either the raw or proportional variant. The results presented here suggest that in many cases, they should be considered as complementary metrics. $\Delta_{raw}$ communicates the absolute size of the conversational change and gives a better indication of volume of relevant messages being posted, whereas $\Delta_{prop}$ is designed to account for the variable activity rates common in social media and allow for effective comparisons between messages released at different times and in different contexts. Each of these variants has shortcomings at the extremes of the activity distribution, however: $\Delta_{raw}$ can be misleading when the baseline activity changes rapidly or press releases follow major events (as highlighted in Figure~\ref{fig:diff_window_echo_examples}); whereas $\Delta_{prop}$ is susceptible to periods of low baseline activity, during which time any increase is proportionally larger. While the level of activity we see in the English language Twitter discourse of climate change means we do not observe this latter challenge in this instance, future applications of $\Delta_{prop}$ to smaller topics or audiences should be aware of this behaviour. Considering both $\Delta_{raw}$ and $\Delta_{prop}$ together provides valuable context and clarification.

In this work, we have presented a new tool with value to communications scholars and particularly those with an interest in agenda setting. Our $\Delta$ metrics help to improve understanding of the influence of specific messages on the public discourse and presents a quantification to aid in the identification of successful strategies. Communications professionals are also likely to benefit from better evaluation of the `echo' of their communications and robust quantification of reception. It should be stressed here however that our metrics should not be used to make causal inference. In the context of our conceptual framework we can consider a null result of small $\Delta$ as indicative of no influence, whereas a large $\Delta$ could arise from a number of (typically impossible to measure and distinguish) confounding factors. When such characterisations are important, we recommend using the $\Delta$ values to identify promising candidates for subsequent qualitative analysis.

This work suggests several promising avenues for future research. The calculation of messaging echo is agnostic to the text domains compared, and could be readily applied to study influence on political conversations or news media. Varying the reference texts may help identify sources of particular messaging and illustrate the path of influence from source to wider usage. Comparing the echo of the same texts across a range of such sources would provide another valuable dimension to this analysis. Being able to differentiate the uptake of strategic communications across target groups will enable more granular analyses and answers to questions such as the extent to which messages reach outside of already engaged individuals.

While our experiments show that sentence representations are of good quality for our purposes (as discussed in the context of our set of labelled text pairs), there may still be bias encoded in the text representations of the model. For example, language containing gender and racial bias has been shown to exist in text corpora commonly used to train large language models~\cite{ferrer_discovering_2020, field_survey_2021}. While ``numerous debiasing techniques [have been] proposed''~\cite{joniak_gender_2022}, evaluating and adapting such a method to suitably fine-tune our model was beyond the scope of our study. However, the modular nature of our method means future work could include the use of a debiased language model or those that are specifically tuned for particular corpora.

Overall, the results presented here demonstrate the value of considering semantic similarity as a means of tracking content alignment between communication mediums. Further investigation of the methods we introduce here presents exciting opportunities for communications scholars and practitioners for richer analysis of agenda setting at much larger scales and will inform future successful campaigning strategies.

\bibliography{refs}

\end{document}